# Fermi-crossing Type-II Dirac fermions and topological surface states in NiTe$_2$

*Saumya Mukherjee[1, 2], Sung Won Jung[1], Sophie F. Weber[3,4], Chunqiang Xu[5], Dong Qian[6], Xiaofeng Xu[5], Pabitra K. Biswas[7], Timur K. Kim[1], Laurent C. Chapon[1], Matthew D. Watson[1], Jeffrey B. Neaton[3,4,8], Cephise Cacho[1]*

[1]Diamond Light Source, Oxfordshire OX11 0DE, United Kingdom.

[2]Clarendon Laboratory, Department of Physics, University of Oxford, Parks Road, Oxford OX1 3PU, United Kingdom

[3]Department of Physics, University of California, Berkeley, California 94720, USA

[4]Molecular Foundry, Lawrence Berkeley National Laboratory, Berkeley, California 94720, USA

[5]Department of Physics, Changshu Institute of Technology, Changshu 215500, China

[6]School of Physics and Astronomy, Shanghai Jiao Tong University, Shanghai 200240, China

[7]ISIS facility, STFC Rutherford Appleton Laboratory, Harwell Science and Innovation Campus, Oxfordshire, OX11 0QX, United Kingdom

[8]Kavli Energy Nanosciences Institute at Berkeley, CA, 94720, USA

Abstract:

Transition-metal dichalcogenides (TMDs) offer an ideal platform to experimentally realize Dirac fermions. However, typically these exotic quasiparticles are located far away from the Fermi level, limiting the contribution of Dirac-like carriers to the transport properties. Here we show that NiTe$_2$ hosts both bulk Type-II Dirac points and topological surface states. The underlying mechanism is shared with other TMDs and based on the generic topological character of the Te *p*-orbital manifold. However, unique to NiTe$_2$, a significant contribution of Ni *d* orbital states shifts the energy of the Type-II Dirac point close to the Fermi level. In addition, one of the topological surface states intersects the Fermi energy and exhibits a remarkably large spin splitting of 120 meV. Our results establish NiTe$_2$ as an exciting candidate for next-generation spintronics devices.

Introduction:

Topologically protected electronic states hold great promise for the development of next-generation electronic and spintronic devices [1, 2, 3]. For topological insulators (TI), such as the well-known case of $Bi_2Se_3$ [4], the carriers associated with the topologically-protected surface states have the unique properties of spin-momentum locking and suppression of backscattering, making the surfaces of such TIs highly attractive platform for spin-transport applications. In the case of topological semimetals, topologically protected crossings occur in the bulk band structure, leading to the formation of bulk Dirac points. In the vicinity of these points, the electronic bands cross linearly, opening up the potential for carriers with ultrahigh-mobility, such as are found in $Cd_3As_2$ [5].

The class of transition metal dichalcogenides (TMDs), already well-known for hosting an array of correlated electronic phenomena, has recently been identified as a materials system which also hosts topologically-protected states [6-11]. Density functional theory (DFT) calculations for a wide range of compounds predict a set of bulk Dirac points and band inversions (associated with topological surface states, TSS), as the chalcogen orbital symmetries combined with a natural hierarchy of energy scales ensure that these states exist rather generically. However, the energies at which they occur is not guaranteed by any physical constraint, and many are predicted to exist in the unoccupied states, where they are hard to verify experimentally. On the other hand, topological states below the Fermi level ($E_F$) can be readily verified and scrutinised by ARPES and spin-ARPES. When these states lie at high binding energies however, they do not affect the transport properties, limiting any potential applications.

Given that TMDs are being increasingly incorporated into device structures [12-17], it is highly desirable to identify a candidate material whose transport properties might be predominantly derived from topologically protected states. An intriguing candidate in this regard is $NiTe_2$. Recent transport measurements have confirmed a substantial magnetoresistance effect, the response becoming large and linear in magnetic field, a feature characteristic of topological semimetals [18]. The Berry phase of the observed quantum oscillations is close to $\pi$, often interpreted as a signature of the contribution of a topologically non-trivial state. Moreover, DFT calculations have suggested that one of the bulk Dirac points of $NiTe_2$ may indeed lie much closer to the Fermi level than in other members of the family where the equivalent states are much further away and thus less relevant to transport (or superconductivity) [6]. These calculations strongly motivate an experimental survey of the electronic structure of $NiTe_2$ using ARPES.

In this paper, we demonstrate with a combination of ARPES and comprehensive DFT calculations that the band structure of $NiTe_2$ shows the formation of type-II Dirac fermions near the Fermi level and a set of spin polarised topological surface states. A Fermi surface consisting of electron pockets associated with type-II Dirac fermions and topological surface states is predicted. The band crossing leading to type-II Dirac fermions is dominated by Te $p$-orbitals, consistent with the mechanism of band inversion from the chalcogen $p$-orbital manifold shown in other TMDs [6, 7]. Our ARPES study shows the formation of electron pockets hosting Dirac carriers at the Fermi level. Topological surface states observed through ARPES agree well with our DFT calculations. One of the surface states is found to intersect the Fermi level, forming electron pockets. The measured Fermi surface map matches well with our calculations and the observation of electron pockets implies finite contribution of Dirac and surface carriers to the transport properties.

Results and discussion:

In what follows, we adopt the tight-binding-based analysis introduced in Refs [6, 7] for understanding the band structure of $NiTe_2$, which is distinct from previously known cases of isovalent and isostructural transition-metal dichalcogenides $TMX_2$ (TM = Pd, Pt; X = Te, Se). Our tight-binding model incorporates two chalcogen sites (Fig 1a), and captures the manner in which the chalcogenide $p$ orbital manifold develops into dispersive bands which generically possess topological characteristics: bulk Dirac points, inverted band gaps (IBGs), and topologically protected surface states (TSS).

In $NiTe_2$, the triply degenerate energy levels of the Te $p_{x,y,z}$-orbitals split into $p_z$ (upper-state) and $p_{x,y}$ (lower-state) manifolds due to crystal-field splitting (CFS) (Fig. 1c). Additionally, spin-orbit coupling (SOC) causes the $p_{x,y}$ orbitals to split into singlets ($R_{4'}$ and $R_{5,6}$ levels) and modifies the energetic separation between $p_z$ and $p_{x,y}$. The $p_z$ orbital transform according to the irreducible representation (IREP) $R_4$ [7]. The two Te sites in the unit cell imply that the $p$-orbitals are split into bonding (B) and antibonding (AB) levels at the $\Gamma$ ($\mathbf{k} = (0, 0, 0)$) and the A ($\mathbf{k} = (0, 0, \pi/c)$)-points of the Brillouin zone [7, 10]. Symmetry wise, the $p$-bands associated with B and AB levels at $\Gamma$ and A points transform as the IREPs of the double space groups $\bar{\Gamma}_8$ ($\bar{A}_8$),$\bar{\Gamma}_4\bar{\Gamma}_5(\bar{A}_4\bar{A}_5)$ and $\bar{\Gamma}_9$ ($\bar{A}_9$) [19, 20]. Because of the phase introduced, the B-AB splitting of the $p_z$-states decreases from the $\Gamma$ to the A point [13]. This induces strong $k_z$ dispersion and a large bandwidth for $p_z$-derived bands as compared to planar $p_{x,y}$ orbitals. For TMDs, this approximation is valid considering the expected large inter-layer-hopping ($t_3$, $t_4$ in Fig. 1a) along the c-axis for $p_z$ states as compared to the $p_{x,y}$ states.

When the bandwidth of $p_z$-derived states become greater than the combination of CFS and SOC, a crossing between the $p_z$ and $p_{x/y}$ states occurs as a function of $k_z$, resulting in band inversion. The states formed due to the B-AB splitting of $p$-orbitals pick up even (+) or odd (-) parity since the Te-atom sites at the $\Gamma$ and A points are located across the crystal inversion point. Therefore, $k_z$-dispersion of a single orbital ($p$) manifold leads to band parity inversion. Fig. 1c, shows the symmetry correspondence between IREPs of the double group at the $\Gamma$ and A point, and along the $\Delta$ symmetry line for all the $k = (0, 0, k_z)$ points with $0 < k_z < \pi/c$. A crossing between $R_{4'}$ ($\pm$) and $R_4$ ($\mp$) causes hybridisation as seen from the identical IREP, $\bar{\Delta}_6$, for these two levels along the $\Delta$ point. This leads to a parity inverted band gap (IBG) (See Fig. 1c). The band gap formed by the bands of opposite parity are capable of hosting topological surface states. On the other hand, $R_4$ ($\pm$) and $R_{5,6}$ ($\mp$) have different symmetry properties along the $\Delta$ line, as they transform differently under 3-fold rotation. This allows band crossing between $R_4$ and $R_{5,6}$ and leads to the formation of the bulk Dirac points (BDPs) along the $k_z$ direction (valid for $0 < k_z < \pi/c$). The location of the BDP in $k$-space is determined by the band width of $R_4$ ($\pm$)-derived bands and the strength of CFS.

The discussion above is based on symmetry, yielding a generic and phenomenological viewpoint. To obtain a more quantitative prediction of the location of the expected bulk Dirac points and IBGs, we turn to ab-initio DFT calculations. Our calculated band structure showing band dispersion and the orbital character of bands is shown in Fig. 2a. The density of states (DoS) confirms that the bands close to the Fermi level, $E_F$, are dominated by Te $p$-orbitals. Ni $d$-orbital derived bands appear away from the Fermi level at $E - E_F \sim -2$ eV (Fig. 2b). The $p_{x/y}$ bands show significantly stronger dispersion along the in-plane high symmetry directions Γ(A)-M (L) and Γ(A)-M (K), compared to the out-of-plane Γ-A direction. However, consistent with the tight-binding analysis, the $p_z$ bands have a large dispersion along Γ-A with wide band width and cross the $p_{x/y}$ bands as a function of $k_z$.

Our DFT calculations indicate several important differences between NiTe$_2$ and the other TMX$_2$ compounds. Weakly dispersive bands of predominantly $d$-character are found at $E – E_F \sim -2$ eV (Fig. 2a, c). The location of these bands is almost 1.5 eV closer to the Fermi level compared to PdTe$_2$, a significant difference on replacing Pd with Ni. In addition, at K close to the Fermi level ($E-E_F \sim 0$ to $-250$ meV), the electron pockets are predominantly formed by bands derived from $d$-orbitals.

The large shift of TM-derived bands towards the Fermi level in NiTe$_2$ increases the hybridisation between Ni-$d$ and the Te-$p$ bands. For example, the crossing of Te $p_z$ and $p_{x,y}$ bands occurs near the A point around $E-E_F \sim -1$ eV, but the $p_z$ bands in this energy range show a significant hybridisation with Ni $d$-orbitals, unlike in other TMX$_2$ compounds. However, this hybridisation does not disrupt the universal mechanism of formation of the inverted band gaps, Dirac fermions and topological surface states in TMX$_2$. A type-I bulk Dirac point (BDP-I) appears at $E-E_F \sim 1.5$ eV, which is followed by a type-II protected bulk Dirac point (BDP-II) at $E-E_F \sim 76$ meV and $k_z = \pm 0.36c^*$ ($c^* = 2\pi/c$). The bands forming the BDP-II are labelled as 1 and 2 (Fig. 2d). These bands cross $E_F$ and form electron pockets. This type-II Dirac fermion has been found in other TMX$_2$; however, the proximity of BDP-II to the Fermi level is unique to NiTe$_2$ [6-10, 18]. In NiTe$_2$, the location of BDP-II in momentum space is slightly closer to A-point than PtTe$_2$ ($k_z = \pm 0.346c^*$) but further away from A than in PdTe$_2$ ($k_z = \pm 0.40c^*$) [7, 9, 10, 21]. This implies that the strength of CFS and bandwidth of $R_4(\pm)$ derived bands is intermediate between PtTe$_2$ and PdTe$_2$. Close to BDP-II, the $p_z$ derived bands form inverted band gap IBG-I at $E-E_F \sim -0.65$ eV (smaller gap $\sim 200$ meV), which is accompanied by IBG-II at $E-E_F \sim -1.5$ eV (larger gap $\sim 1$ eV). We want to point out that the inverted band gaps (IBGs) at the A-point have inverted parity, implying the existence of topologically ordered states.

With these predictions in hand, we now turn to the experimental measurements of the bulk and surface electronic structure of NiTe$_2$. First, we focus on tracking the BDP-II and surface states along Γ-A using photon-energy-dependent ARPES and measure the occupied states below the Fermi level (Fig. 2e). The band features are broadened due to the finite $k_z$ resolution of photoemission but the states are in good agreement with calculated bulk bands. Both IBGs, near Fermi level ($E-E_F \sim -0.65$ eV) and away from it ($E-E_F \sim -1.5$ eV) are found. The BDP-II lying above the Fermi level is not accessible by ARPES, but the bands 1 and 2 forming the BDP-II are observed at the A-point ($k_z = \pi/c$). Interestingly, a non-dispersive two-dimensional band is found at $E-E_F \sim -1.35$ eV close to A-point which does not match any calculated bulk band (marked in Fig. 2e).

To better understand the band structure, we compared the experimental and calculated in-plane band dispersion along A-L and A-H in Fig. 3. The bulk bands show good agreement with bulk DFT calculations (Fig. 3a, b, d, e). Our DFT slab calculations (see Methods for details) show a set of surface states which match well with the experimental data (Fig. 3 c,f). The non-dispersive feature at E-$E_F$ ~ -1.35 eV is reproduced as one of the surface states by the slab calculations. We assign the surface state as TSS2. At the A-point, TSS2 forms sharp and intense bands. This feature threads through the IBG-II and connects the time-reversal invariant momentum (TRIM) points (here A, L, H). TSS2 represents a topological non-trivial band but lies far away from the Fermi level.

Slightly closer to the Fermi level, our DFT calculations show a topological surface state at E-$E_F$ ~ - 0.65 eV within the band gap IBG-I. This is named as TSS1 and lies within the manifold of bulk bands. TSS1 experiences strong interaction with the bulk states and is classified as a surface resonance state. In ARPES spectra, we observe weak spectral intensity of TSS1 at A, which matches well with the calculations (Fig. 3a, c). However, the mixing of TSS1 with bulk bands makes it difficult to resolve.

At the Fermi level, a surface state is visible in our DFT surface band structure along A-L (Fig. 3c). This is assigned as TSS0, which shows a large spin splitting of around 120 meV and lies between E-$E_F$ ~ 0 **to** - 200 meV. TSS0 is not present along A-H (Fig. 3f). Experimentally, TSS0 is found between A and L with in-plane momentum $k_{//}$ ~ 0.5 (Å)$^{-1}$. TSS0 is formed from two parabolic-like upper (labelled as $\varepsilon$) and lower ($\gamma$) branches which intersect EF and form electron pockets (see Fig. 4a).

The $k_z$ – dependence of the electronic states spanning over a wide photon energy range ($hv$ = 20 -120 eV) shows that TSS0 is non-dispersive along Γ-A (Fig. 4b). This confirms that TSS0 is two-dimensional. The vicinity of TSS0 to EF in NiTe$_2$ makes it unique compared to other surface states and implies finite contribution of topological surface carriers to the non-trivial transport properties. All topological surface states found previously in other TMX$_2$ are also found in NiTe$_2$ [6-10], but notably TSS0 is absent in PtSe$_2$.

To complete our studies, we map the Fermi surface of NiTe$_2$ using photon energy tuned to the A-point ($hv$ = 23 eV) (Fig. 3b). It is found that the Fermi surface is formed by TSS0 and the bulk bands crossing the Fermi level. The $\varepsilon$ and $\gamma$ bands of TSS0 are identified. These bands form electron pockets at the Fermi level. The pockets formed by $\varepsilon$ are more circular than the pockets of the $\gamma$ band. Identical to PdTe$_2$, these bands form arc-like features and imply small projected bulk band gaps. At the A-point, band 2 forms an electron pocket (labelled as $\alpha$) and the projected bulk bands form a hexagonal shape. We show that the calculated bulk Fermi surface agrees well with the experimental data (see Inset of Fig. 3b).

To identify the topological character of the surface states, the associated spin texture is calculated as shown in Fig. 3c, d. In our description, the *x*- and *y*-axis are along A-L and A-H, respectively. TSS2 has a chiral spin texture in the *x-y* plane with finite <$S_x$> and <$S_y$> components. Negligible <$S_z$> component is found (see Supplementary Information). The strong interaction with bulk bands for TSS1 limits the estimation of spin polarization, and we do not include polarization for TSS1. The electron pockets formed by TSS0 along L-A-L (indicated in Fig. 4c) are well separated from the bulk and hence it is possible to resolve spin texture here. We find that TSS0 has non-zero spin component solely along the *y*-direction <$S_y$>, which is normal to the A-L direction. The branches of TSS0, $\varepsilon$ and $\gamma$, show opposite spin polarization.

We want to point out that the spin-split branches of the surface state TSS0 close to the Fermi level are connected to the conduction band and the valence band separately, which highlights the topological non-trivial character [22]. For completeness, we include spin polarization for TSS0 along the entire L'-A-L and H'-A-H paths, but note that the only location where the surface character has not significantly hybridized with bulk is at the electron pockets, so calculated spin texture at all other locations does not accurately represent surface state spin texture of TSS0.

Conclusion

We have shown that NiTe2 exhibits topological surface states and bulk type-II Dirac points, which are derived solely from the Te $5p$ orbitals, consistent with the generic band inversion mechanism for transition metal dichalcogenides. However, the choice of Ni as the transition metal leads to 3d states significantly closer to the Fermi level compared with (Pd, Pt)(Se, Te)$_2$, allowing additional d-p hybridisations which tune the bulk Dirac point very close to EF. The band dispersions away from this BDP-II, which form the alpha electron pocket, thus have a topologically non-trivial character, likely to explain the non-trivial Berry phase of a small electron-like pocket observed in transport studies [18]. We have also shown that NiTe$_2$ harbours a unique topological surface state, TSS0, with one of the largest spin-splitting of up to 120 meV reported for any Fermi-crossing surface state of a transition metal dichalcogenide. For future studies, an exciting approach would be to exfoliate individual monolayers or grow very thin films of NiTe$_2$, with a significantly increased contribution of these topological surface carriers to the transport properties. Thus both the surface and bulk electronic structures of this material are candidates for engineering novel spintronics devices underpinned by a robust and generic topological mechanism.

Methods:

**ARPES measurements**

High quality single crystals of $NiTe_2$ were grown by chemical vapour transport as discussed elsewhere [18]. We used the high-resolution vacuum ultraviolet (VUV) ARPES branch of I05 beam line at Diamond Light Source, UK [23]. The samples were cleaved at 10 K, in ultra-high vacuum chamber of I05. Linear horizontal and vertical polarization of photon at energies between hν = 20 and 120 eV is used. The inner potential $V_0$ = 16 eV of $NiTe_2$ is determined from the $k_z$ dispersion from photon-energy-dependent ARPES studies using a free electron final state model.

**First principle calculations**

Our calculations were performed within density functional theory (DFT) with room-temperature lattice parameters (details on lattice parameters is discussed elsewhere [18]). We employ the Vienna *ab-initio* simulation package (VASP) [24] with generalized gradient approximation (GGA) using the Perdew-Burke-Ernzerhof (PBE) functional [25] and projector augmented-wave (PAW) method [26]. We treat 3*p*, 3*d*, and 4*s*, and 5*s* and 5*p* electrons as valence for Ni and Te, respectively. We use an energy cut off of 550 eV for our plane-wave basis set, with a Γ-centred *k*-point mesh of 14x14x14 for the primitive unit cell. For surface state features, we perform full self-consistent DFT calculations using a Te-terminated supercell with a slab of ten primitive unit cells of $NiTe_2$ stacked along the [001] direction, with 15 Å of vacuum.

**Data availability**

The datasets that support the findings of this study are available from the corresponding author upon reasonable request.

References:

[1] Naoto Nagaosa. A New State of Quantum Matter. *Science.* **318**, 5851, 758-759 (2007).

[2] Moore, J. Topological insulators: The next generation. *Nat. Phys.* **5**, 378–380 (2009).

[3] Brumfiel, Geoff. Topological insulators:Star material. *Nature* **466,** 310-311 (2010).

[4] Hasan, M. Z. and Kane, C. L. Colloquium: Topological insulators. *Rev. Mod. Phys.* **82,** 3045-3067 (2010).

[5] Madhab Neupane, Su-Yang Xu, Raman Sankar, Nasser Alidoust, Guang Bian, Chang Liu, Ilya Belopolski, Tay-Rong Chang, Horng-Tay Jeng, Hsin Lin, Arun Bansil, Fangcheng Chou & M. Zahid Hasan: Observation of a three-dimensional topological Dirac semimetal phase in high-mobility Cd3As2. *Nature Communications* **5,** 3786 (2014).

[6] O. J. Clark, M. J. Neat, K. Okawa, L. Bawden, I. Marković, F. Mazzola, J. Feng, V. Sunko, J. M. Riley, W. Meevasana, J. Fujii, I. Vobornik, T. K. Kim, M. Hoesch, T. Sasagawa, P. Wahl, M. S. Bahramy, and P. D. C. King: Fermiology and Superconductivity of Topological Surface States in PdTe2. *Phys. Rev. Lett.* **120**, 156401 (2018).

[7] M. S. Bahramy, O. J. Clark, B.-J. Yang, J. Feng, L. Bawden, J. M. Riley, I. Marković, F. Mazzola, V. Sunko, D. Biswas, S. P. Cooil, M. Jorge, J. W. Wells, M. Leandersson, T. Balasubramanian, J. Fujii, I. Vobornik, J. E. Rault, T. K. Kim, M. Hoesch, K. Okawa, M. Asakawa, T. Sasagawa, T. Eknapakul, W. Meevasana & P. D. C. King: Ubiquitous formation of bulk Dirac cones and topological surface states from a single orbital manifold in transition-metal dichalcogenides. *Nature Materials* **17**, 21–28 (2018).

[8] Huaqing Huang, Shuyun Zhou, and Wenhui Duan : Type-II Dirac fermions in the PtSe2 class of transition metal dichalcogenides. *Phys. Rev. B* **94**, 121117 (R) (2016).

[9] Mingzhe Yan, Huaqing Huang, Kenan Zhang, Eryin Wang, Wei Yao, Ke Deng, Guoliang Wan, Hongyun Zhang, Masashi Arita, Haitao Yang, Zhe Sun, Hong Yao, Yang Wu, Shoushan Fan, Wenhui Duan & Shuyun Zhou: Lorentz-violating type-II Dirac fermions in transition metal dichalcogenide PtTe2. *Nature Communications* **8**, 257 (2017).

[10] O J Clark, F Mazzola, I Marković, J M Riley, J Feng, B-J Yang, K Sumida, T Okuda, J Fujii, I Vobornik, T K Kim, K Okawa, T Sasagawa, M S Bahramy and P D C King: A general route to form topologically-protected surface and bulk Dirac fermions along high-symmetry lines. *Electronic Structure* **1,** 1, (2019).

[11] Sajedeh Manzeli, Dmitry Ovchinnikov, Diego Pasquier, Oleg V. Yazyev & Andras Kis: 2D transition metal dichalcogenides. *Nature Review Materials* **2**, 17033 (2017).

[12] Wenchao Tian, Wenbo Yu, Jing Shi and Yongkun Wang. The Property, Preparation and Application of Topological Insulators: A Review. *Materials* **10**, 814, (2017).

[13] Teweldebrhan, D.; Balandin, A.A. "Graphene-Like" exfoliation of atomically-thin films of Bi2Te3 and related materials: Applications in thermoelectrics and topological insulators. *ECS Trans.* **33**, 103–117, (2010).

[14] Ganesh R. Bhimanapati, Zhong Lin, Vincent Meunier, Yeonwoong Jung, Judy Cha, Saptarshi Das, Di Xiao, Youngwoo Son, Michael S. Strano, Valentino R. Cooper, Liangbo Liang, Steven G. Louie, Emilie Ringe, Wu Zhou, O Steve S. Kim, Rajesh R. Naik, Bobby G. Sumpter, O Humberto Terrones, Fengnian Xia, Yeliang Wang, Jun Zhu, Deji Akinwande, Nasim Alem, Jon A. Schuller, Raymond E. Schaak, Mauricio Terrones, and Joshua A. Robinson. Recent Advances in Two-Dimensional Materials beyond Graphene. *ACS Nano* **9**, 12, 11509-11539 (2015).

[15] B. Radisavljevic, A. Radenovic, J. Brivio1, V. Giacometti1 and A. Kis. Single-layer MoS2 transistors. *Nature Nanotechnology* **6**, 147–150 (2011).

[16] Y. J. Zhang1, T. Oka, R. Suzuki, J. T. Ye, Y. Iwasa. Electrically Switchable Chiral Light-Emitting Transistor. *Science* 344, 6185, 725-728, (2014).

[17] Wonbong Choi, Nitin Choudhary, Gang Hee Han Juhong Park, Deji Akinwande, Young Hee Lee. Recent development of two-dimensional transition metal dichalcogenides and their applications. *Materials Today* 20, **3** (2017).

[18] Chunqiang Xu, Bin Li, Wenhe Jiao, Wei Zhou, Bin Qian, Raman Sankar, Nikolai D. Zhigadlo, Yanpeng Qi, Dong Qian, Fang-Cheng Chou, Xiaofeng Xu: Topological Type-II Dirac Fermions Approaching the Fermi Level in a Transition Metal Dichalcogenide NiTe2. *Chem. Mater.* **30,** 14, 4823-4830, (2018).

[19] M. I. Aroyo, J. M. Perez-Mato, C. Capillas, E. Kroumova, S. Ivantchev, G. Madariaga, A. Kirov & H. Wondratschek. "Bilbao Crystallographic Server I: Databases and crystallographic computing programs". *Zeitschrift fuer Kristallographie* (2006), **221**, 1, 15-27.

[20] M. I. Aroyo, A. Kirov, C. Capillas, J. M. Perez-Mato & H. Wondratschek. "Bilbao Crystallographic Server II: Representations of crystallographic point groups and space groups". Acta Cryst. (2006), A62, 115-128.

[21] Fucong Fei, Xiangyan Bo, Rui Wang, Bin Wu, Juan Jiang, Dongzhi Fu, Ming Gao, Hao Zheng, Yulin Chen, Xuefeng Wang, Haijun Bu, Fengqi Song, Xiangang Wan, Baigeng Wang, and Guanghou Wang: Nontrivial Berry phase and type-II Dirac transport in the layered material PdTe2. *Phys. Rev. B* **96**, 041201 (**R**) (2017).

[22] G. Bian, T. Miller, and T.-C. Chiang. Passage from Spin-Polarized Surface States to Unpolarized Quantum Well States in Topologically Nontrivial Sb Films. *Physical Review Letters* **107**, 036802 (2011).

[23] M. Hoesch, T. K. Kim, P. Dudin, H. Wang, S. Scott, P. Harris, S. Patel, M. Matthews, D. Hawkins, S. G. Alcock, T. Richter, J. J. Mudd, M. Basham, L. Pratt, P. Leicester, E. C. Longhi1, A. Tamai, and F. Baumberger: A facility for the analysis of the electronic structures of solids and their surfaces by synchrotron radiation photoelectron spectroscopy. Review of Scientific Instruments **88**, 013106 (2017).

[24] G. Kresse and J. Furthmüller: Efficient iterative schemes for ab initio total-energy calculations using a plane-wave basis set. *Phys. Rev. B* **54**, 11169 (1996).

[25] John P. Perdew and Wang Yue: Accurate and simple density functional for the electronic exchange energy: Generalized gradient approximation. *Phys. Rev. B* **33**, 8800(**R**) (1986); Erratum *Phys. Rev. B* **40**, 3399 (1989).

[26] P. E. Blöchl: Projector augmented-wave method. *Phys. Rev. B* **50**, 17953 (1994).

[27] Momma, K. & Izumi, F. VESTA 3 for three-dimensional visualization of crystal, volumetric and morphology data. *Journal of Applied Crystallography* **44**, 1272–1276, (2011).

**Acknowledgements**

We would like to thank P. D. C. King and Pavel Dudin for a fruitful discussion. This work is based on experiments performed at the Diamond Light Source, Didcot, United Kingdom. Data of this work were taken at Diamond Light Source (Beamline I05, proposal No. NT21591). S. Mukherjee acknowledges financial support from the European Unions Horizon 2020 research and innovation programme under the Marie Skodowska-Curie Grant Agreement (GA) No. 665593 awarded to the Science and Technology Facilities Council. S. F. W. and J. B. N. were supported by the Center for Novel Pathways to Quantum Coherence in Materials, an Energy Frontier Research Center funded by the US Department of Energy, Director, Office of Science, Office of Basic Energy Sciences under Contract No. DE-AC02-05CH11231. Computational resources provided by the Berkeley Research Computing (BRC) program. The drawings of crystal structures are produced by VESTA program [27], which is acknowledged.

**Author Contributions**

S. M and S. W. J contributed equally to this work. S.F.W and J. B.N performed the theoretical calculations. The single crystals were prepared by C. X, D. Q and X. X. APRES data were collected by S. M, S. W. J, T. K. K, C. C and analysed by S. W. J and S. M. The overall project planning and direction was led by S. M and S. W. J and supervised by T. K. K, P. K. B, L. C. C and C. C. The manuscript was prepared by S. M, M. D. W, and S. W. J with input from all co-authors.

**Corresponding authors**

Correspondence to Saumya Mukherjee or Cephise Cacho.

**Competing interests**

The authors declare no competing financial or non-financial interests.

Figures:

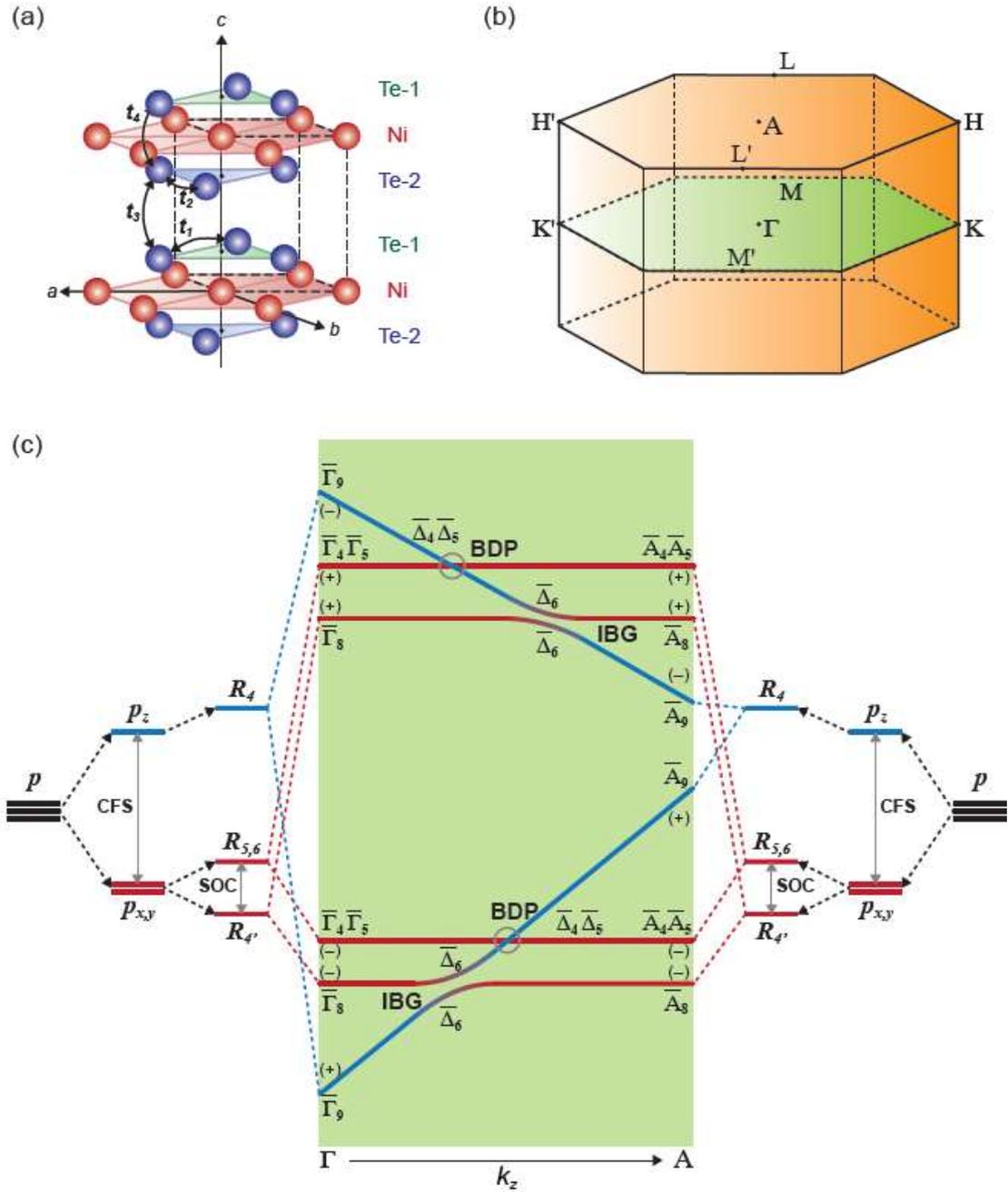

Figure. 1: (a) Crystal structure of 1T-NiTe$_2$ (space group: *P-3m1*), composed of hexagonal basal planes (*ab*-planes) of Ni-atom (red sphere) coordinated to the Ni-atom at the centre and triangular layers with inequivalent Te atomic site (blue sphere) above (Te-1) and below (Te-2) the basal plane along the (001)-direction [18]. The hopping parameters between 2-site Te *p*-orbitals are categorised as intra-layer hopping (*$t_1$ = $t_2$*), interlayer hopping (*$t_3$*) within the unit cell and between two unit cells (*$t_4$*). (b) Brillouin zone. (c) Hierarchy of *p*-orbital derived energy levels at Γ and A-point showing the inverted band gaps (IBG), Dirac points (BDP: circled) and topological states. The symmetry of the states are labelled with IRREPs (Γ$_i$, A$_j$, Δ$_k$) and parity (+/-). Inspired from Ref. [7, 10].

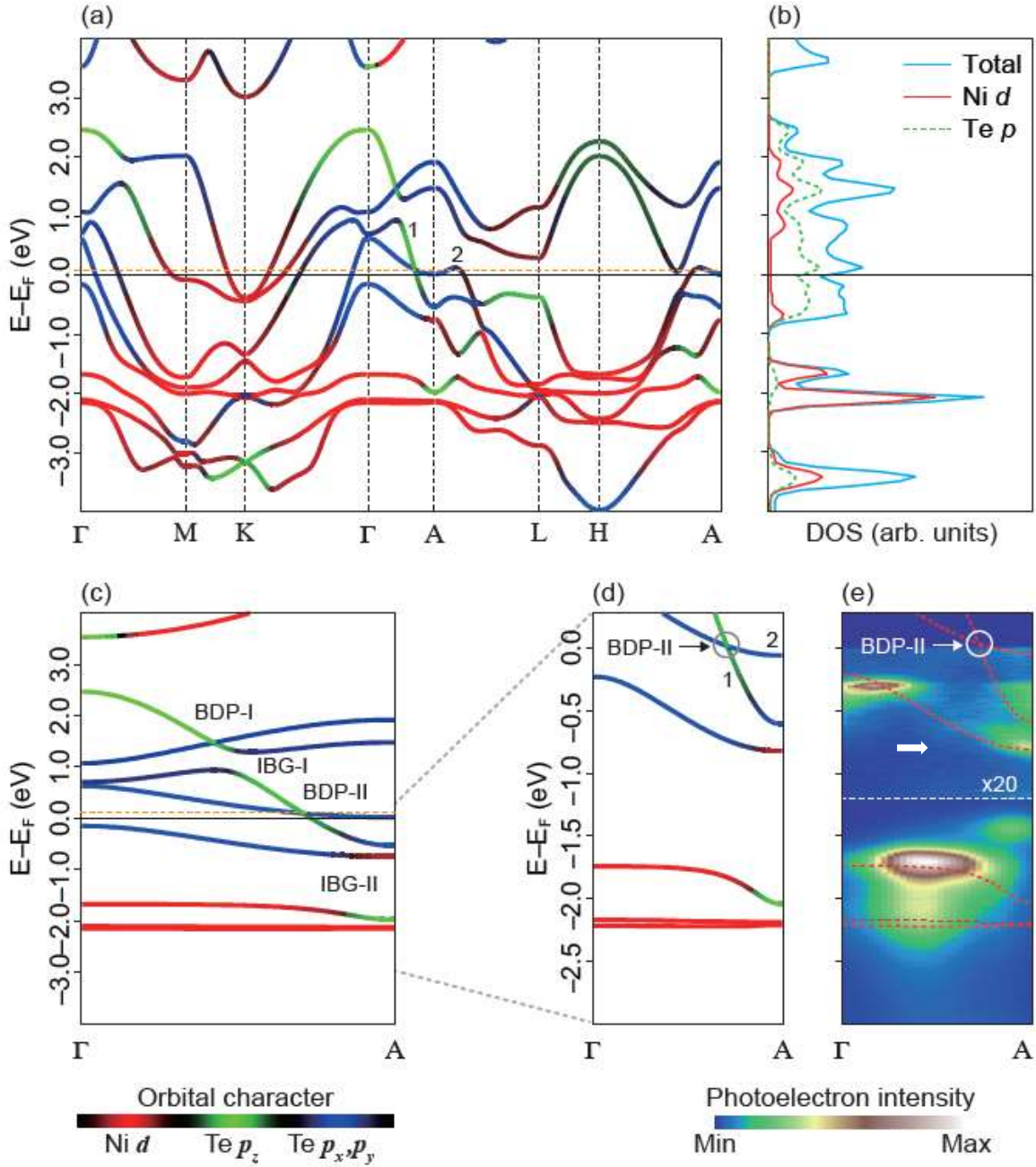

Figure. 2: (a) Electronic DFT bulk band structures with orbital character of bands. Type-II Dirac fermion formed by crossing of band 1 and 2 near Fermi level. (b) Density of states (DoS) showing dominant contribution of Te $p$-bands at the Fermi Level compared to Ni $d$-bands. Above Fermi level, the DoS spectra is scaled up by factor of 2. (c) Band dispersion along the Γ-A direction with inverted band gaps (IBG) and bulk Dirac points (BDP) marked, (d) and (e) Zoomed-in dispersion and measured photon-energy dispersion with linear horizontal polarization along the Γ-A direction. A non-dispersive feature is marked with an arrow, which does not match with bulk DFT calculation.

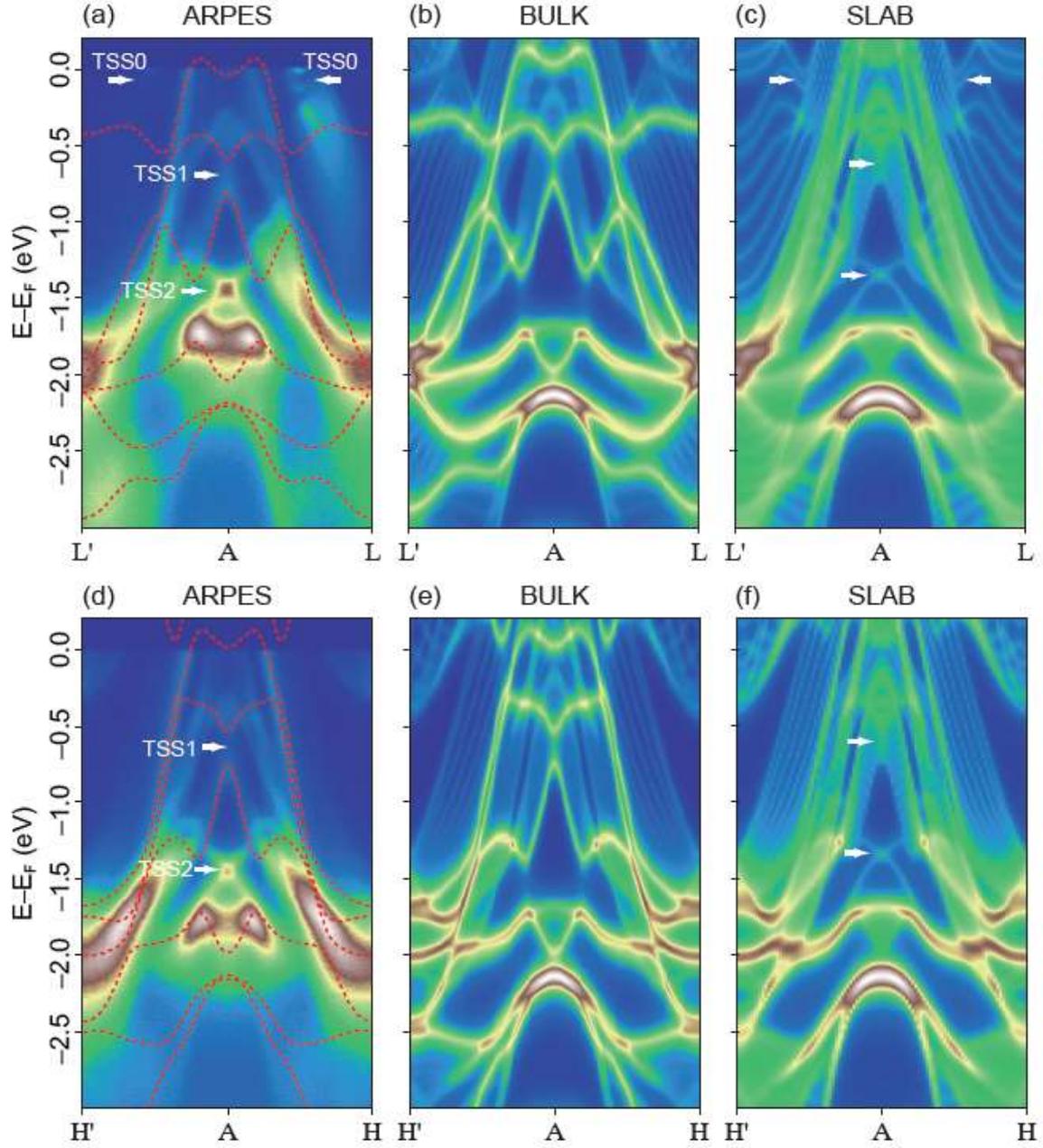

Figure. 3: Spectral band dispersion along in-plane L'-A-L (a-c) and H'-A-H (d-e) direction, probed with photon energy $h\nu = 99$ eV (a) experimental ARPES data with linear horizontal polarization, supercell calculation integrated along $k_z$ with (b) bulk DFT and (c) slab surface. Arrows mark the position of the topological surface states (TSS). The overlaid dotted red lines in (a, d) represent the calculated bulk DFT bands.

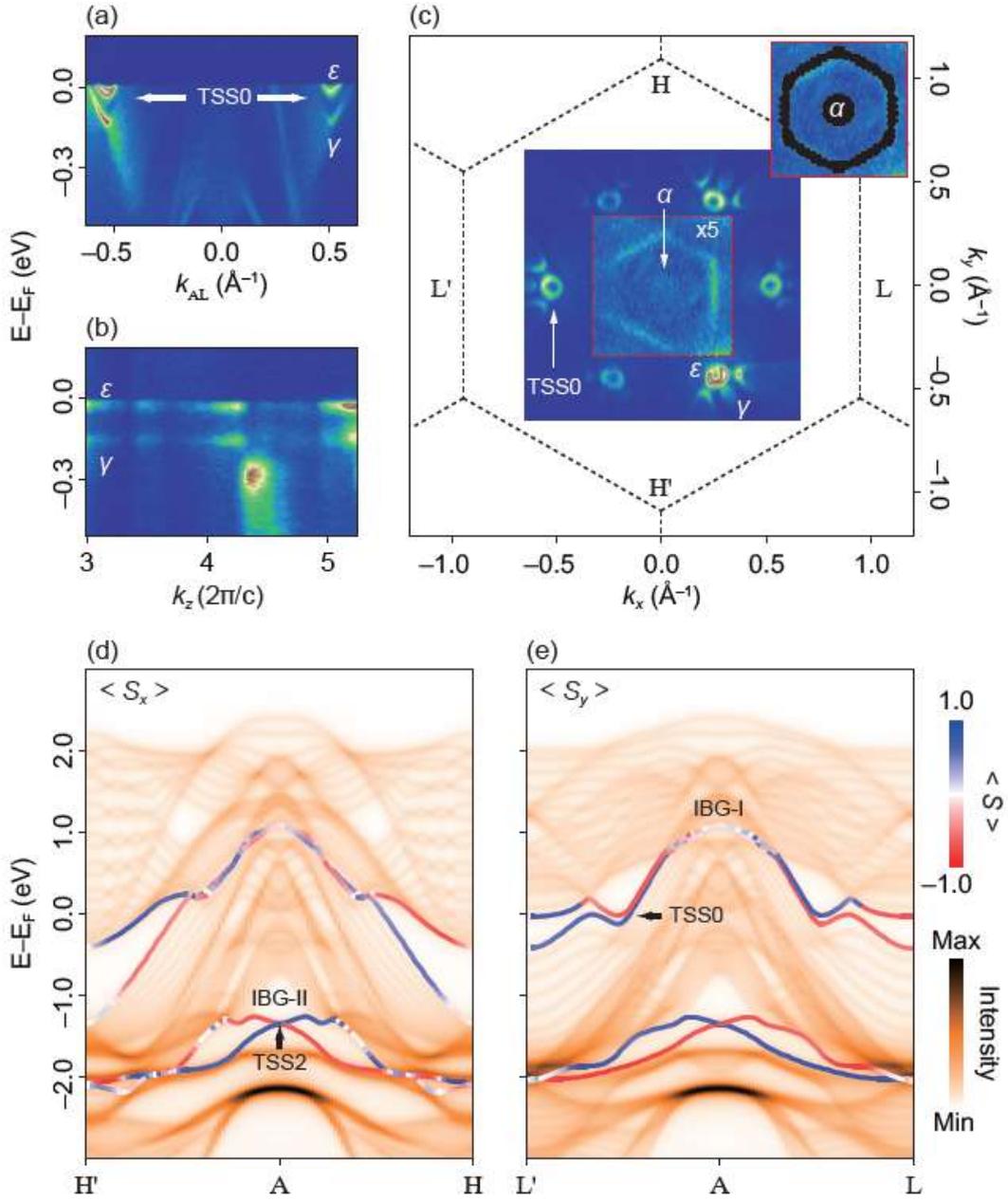

Figure. 4: (a) In-plane dispersion of TSS0 forming electron pockets labelled as $\varepsilon$ and $\gamma$ and (b) the non-dispersive behaviour of TSS0 surface state close to the Fermi level in the photon-energy dependence. (c) Fermi surface measured with horizontal polarization of light and photon energy, $h\nu = 23$ eV, which corresponds to the A-plane. The region under the red dotted box is magnified by a factor of 5 to show the electron pockets $\alpha$ (Band 2), $\varepsilon$ and $\gamma$ (TSS0) and the bulk bands. (Inset: calculated bulk Fermi surface consisting of the circular $\alpha$ electron pocket and bands forming hexagonal contour, in good agreement with measured data). (d) and (e) Slab calculation with projected spin components of the topological surface states. TSS2 shows finite $\langle S_x \rangle$ and $\langle S_y \rangle$ components normal to H'-A-H and L'-A-L direction, respectively. TSS0 has finite spin component $\langle S_y \rangle$ normal to L'-A-L (note that TSS0 is hybridized with bulk in all other regions of the plots, so no further conclusions about surface state spin texture can be made).

*Supplementary Information*

**Fermi-crossing Type-II Dirac fermions and topological surface states in NiTe$_2$**

*Saumya Mukherjee[1,2], Sung Won Jung[1], Sophie F. Weber[3,4], Chunqiang Xu[5], Dong Qian[6], Xiaofeng Xu[5], Pabitra K. Biswas[7], Timur K. Kim[1], Laurent C. Chapon[1], Matthew D. Watson[1], Jeffrey B. Neaton[3,4,8], Cephise Cacho[1]*

[1]Diamond Light Source, Oxfordshire OX11 0DE, United Kingdom.

[2]Clarendon Laboratory, Department of Physics, University of Oxford, Parks Road, Oxford OX1 3PU, United Kingdom

[3]Department of Physics, University of California, Berkeley, California 94720, USA

[4]Molecular Foundry, Lawrence Berkeley National Laboratory, Berkeley, California 94720, USA

[5]Department of Physics, Changshu Institute of Technology, Changshu 215500, China

[6]School of Physics and Astronomy, Shanghai Jiao Tong University, Shanghai 200240, China

[7]ISIS facility, STFC Rutherford Appleton Laboratory, Harwell Science and Innovation Campus, Oxfordshire, OX11 0QX, United Kingdom

[8]Kavli Energy Nanosciences Institute at Berkeley, CA, 94720, USA

In Supplementary information we present the calculated spin texture of TSS0 and TSS2 showing the spin components along *x*, *y* and *z* directions.

Figures:

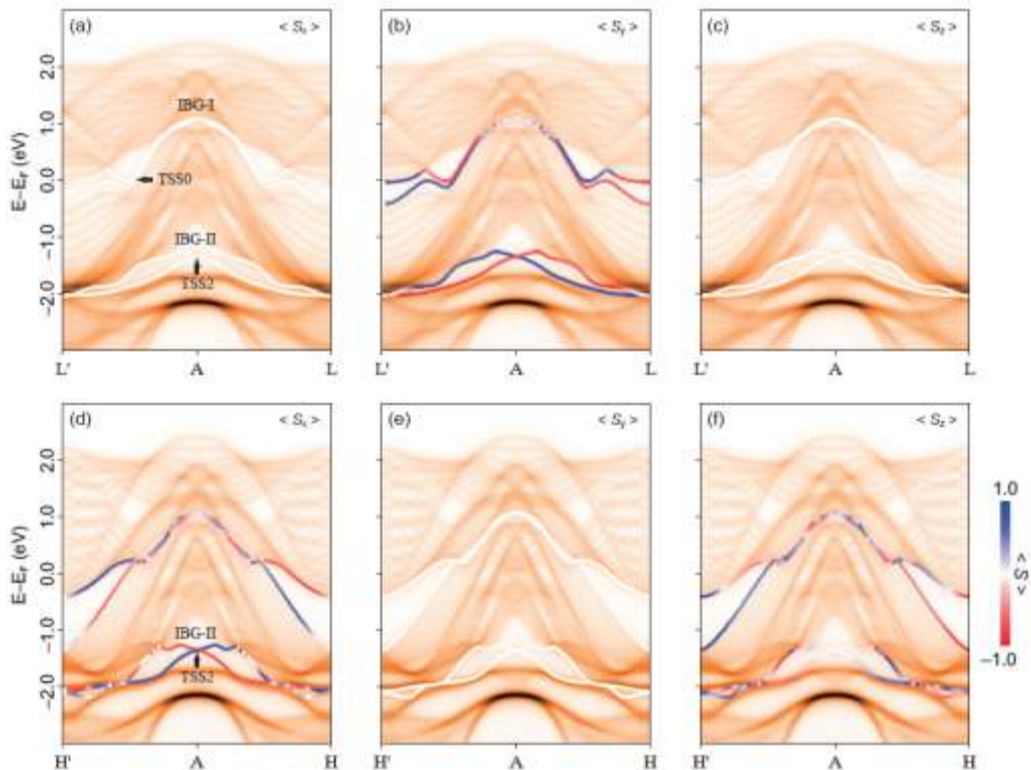

S1: Calculated spin texture of surface states showing spin components $<S_x>$ (a, d), $<S_y>$ (b, e) and $<S_z>$ (c, f).

15